**Cosmological Magnetic Fields: generation during Inflation and Evolution**

Zuxiang Dai

**Supervisor:    Prof. Christoph Schmid**

Diploma thesis in theoretical Physics

*ETH* Zürich
Winter Semester 2001/2002

Cosmological Magnetic Fields: generation during Inflation and Evolution - i -

# CONTENTS

**Abstract**

This paper concerns the generation and evolution of the cosmological (large-scale $\sim Mpc$) magnetic fields in inflationary universe. The universe during inflation is represented by de Sitter space-time.

We started with the Maxwell equations in spatially flat Friedmann-Robertson-Walker (FRW) Cosmologies. Then we calculated the wave equations of the magnetic field and electric field for the evolution. We consider the input current that was produced from a massless charged scalar complex field. This field minimally coupled to both gravity and the electromagnetic fields. The Lagrangian for massless scalar electrodynamics is then $\mathscr{L} = \sqrt{-g}\left(D_\mu \phi (D^\mu \phi)^* - \frac{1}{4} F_{\mu\nu} F^{\mu\nu}\right)$. The complex scalar field $\phi$ couples to electromagnetism through the usual gauge covariant derivative $D_\mu = \partial_\mu - ieA_\mu$. After the quantum field theoretical deduction for the current, we put it back into the wave equation of the magnetic field.

After solving this wave equation, our result is $a^2 B \sim \frac{eH}{\sqrt{2}k^2}\left|\sin\sqrt{2}k\eta\right|$. At the time $\eta_{RH}$ we have $B_{RH} = \frac{e}{k_{phys}}$. This may imply that the breaking of the conformal invariance due to the minimal coupling of a massless charged scalar complex field to both gravitational and electromagnetic fields is not sufficient for the production of seed galactic magnetic fields during inflation. But since we are interested in the large-scale cosmological magnetic field, this could be still a candidate, because of the $1/k$ factor.

# 1   Introduction and Conclusions

Cosmology is the study of the large-scale structure and evolution of the universe. Magnetic fields of galaxies present an interesting problem. The first question is whether the field existed before the galaxy (Hoyle 1958) or was produced in a dynamo in the galaxy (Parker 1975). If the former occurred, magnetic stress could have played an important role in the formation of galaxies.

Our galaxy and many other spiral galaxies are endowed with coherent magnetic fields (ordered on scales $\geq 10$ kpc ) with typical strength $\sim 3 \times 10^{-6}$ G. Today, magnetic fields are present throughout the universe and play an important role in a multitude of astrophysical phenomena. For instance, the magnetic field of our galaxy plays an important role in the dynamics of our galaxy – confining cosmic rays, transferring angular momentum away from protostellar clouds so that they can collapse and become stars (without the loss of angular momentum, protostellar clouds would collapse to a low-density, centrifugally supported, non-starlike state). [5]

Many astrophysicists believe that galactic magnetic fields are generated and maintained by dynamo action (where by the energy associated with the differential rotation of spiral galaxies is converted into magnetic field energy) [5]. The dynamo mechanism is only a means of amplification and dynamos require seed magnetic fields. If a galactic dynamo has operated over the entire age of the galaxy ($\sim 10$ G yr), it could have amplified a seed field by a factor of $e^{[O(30)]} \sim 1 \cdot 10^{13}$, implying that a seed magnetic field $\sim 3 \times 10^{-19}$ G is required [5]. Some astrophysicists believe the galactic magnetic field owes its existence to primeval magnetic flux trapped in the gas that collapsed to form the galaxy [5]; in this case the primeval fields strength required is much greater — at least the field strength that is observed today $\sim 3 \times 10^{-6}$ G .[5]

Since the Universe has been a good conductor throughout most of its history, any primeval, cosmological magnetic field present will evolve conserving magnetic flux: $a^2 B \sim const$ .[5] The magnetic fields produced are then uninterestingly small unless the conformal invariance of the electromagnetic field is broken. An attractive and very economical idea on the possible primordial origin of the galactic magnetic fields was suggested in [5]. That is: a massless charged scalar complex field coupled to both gravity and the electromagnetic fields. In short, it is based on the following observation. While the coupling of electromagnetic field to the metric and to the charged fields is conformally invariant (this is not necessarily true in the models with dynamical dilaton), the coupling of the charged scalar field to gravity is not[13].

Thus classical fluctuations with wavelengths $\geq H^{-1}$ in massless, minimally coupled fields could grow "superadiabatically", i.e., have their energy density ($\simeq k \frac{d\rho}{dk}$ in mode $k$) decrease only as $\sim a^{-2}$, rather than the usual $\sim a^{-4}$ ("adiabatic" result)[5].

The vacuum fluctuations of the charged scalar field could be amplified during inflation at super-horizon scales, leading potentially to non-trivial correlations of the electric currents and charges over cosmological distances. The fluctuations of electric currents, in turn, may induce magnetic fields through Maxwell equations at the corresponding scales[13]. In this paper, we have developed the wave equation of the Magnetic fields.

In short: In the early universe gravity acting on a quantum matter field amplifies energy density fluctuations above the unavoidable vacuum fluctuations. These primordial fluctuations may serve as seeds for the subsequent formation of the observable large-scale structure [16].

No detailed computations were carried out in [5] in order to substantiate this idea.

In this paper, we started with the Maxwell equations in curved space-time. In our case, we write the Maxwell equations in components in spatially flat Friedmann-Robertson-Walker (FRW) space-time $ds^2 = g_{\mu\nu}dx^\mu dx^\nu = dt^2 - a^2(t)\big((dx^1)^2 + (dx^2)^2 + (dx^3)^2\big)$. The normal operations like gradient, curl are changed in curved space-time. We have details in Appendix A.

After we obtain the Maxwell Equations in curved space-time, we developed the wave equations for the magnetic field and the electric field.

For the input current of the wave equation, we have to shift to quantum field theory in curved space-time.

We consider a massless charged scalar field $\phi$ without self-interactions and with minimal coupling to both gravity and the electromagnetic fields under de Sitter background geometry. We notice here, the field needs not to be exactly massless, it is enough if $m \ll H$.

The state of the quantum field is dictated by the inflationary paradigm:

1.) Every mode with comoving wave number $k$ which is of interest to physics observable today had a physical wavelength much smaller than the Hubble radius at early enough times during the inflationary era, i.e. $\lambda_{phys} \ll H^{-1}$. At this early stage of inflation every relevant mode had $k_{phys}^2 \gg R$, thus the curvature $R$ is negligible for the physically relevant modes. [15]

2.) According to the inflationary paradigm we consider the minimal quantum fluctuations, i.e. in every relevant mode we start out with Minkowski vacuum fluctuations. During inflation the initial fluctuations in each mode get amplified by the gravitational tide forces. In the Heisenberg picture of time evolution the states are defined to be time independent and denoted by the properties of the initial state, in our case $|0\ in\rangle$ for all relevant modes (observable today). This state of the quantum fields is the Bunch-Davies state [15] insofar as modes with physical wavelengths today smaller than the present Hubble radius are concerned.

We notice here that we normalize our comoving scales $a$ by setting today (at end of the inflation era and begin of the Reheating era) $a_{RH} \equiv 1$, thus $\lambda_{phys} = a\lambda = \lambda$, therefore at this time the physical and commoving distances coincide and also $k_{phys} = k$. Just as usually, we use here natural units $\hbar = c = k_B = 1$.

During investigate the correlation function $C_\phi(l)$, where $l \equiv |\vec{x} - \vec{x}'|$, we analyzed the energy density fluctuations from this boson field, which with vanishing expectation values, $\langle\phi\rangle = 0$. To avoid the divergence on $l \to 0$, and also since a test body measures the field strength averaged over some region in space, it is more realistic to consider a spatially smeared field operator $F_R \equiv \int d^3x' W_R(|\vec{x} - \vec{x}'|)F(\vec{x}')$ where $W_R$ is the normalized window function with linear smearing scale $R$. Afterwards we computed the correlation function for the current $C_{\vec\nabla \times \hat{j}}(l)$.

After we obtain the current, we try to solve the wave equation of the magnetic field. At first we do Fourier transformation for the magnetic field and the current. Thus we simplified our partial differential equation to an ordinary differential equation. And we calculate this equation based on single modes. A given Fourier component will be labeled by its comoving wavelength $\lambda$ or more conventionally with its comoving wave number $k = \frac{2\pi}{\lambda}$. The most important reason that we use the comoving wave number is, different modes decouple, i.e., the comoving vector $\vec{k}$ is conserved.

After we solve this equation, we obtain the result that $a^2 B \sim \frac{eH}{\sqrt{2}k^2}\left|\sin\sqrt{2}k\eta\right|$. We see that the magnetic field is proportional to wave length, so the smaller the wave number the bigger wavelength, and the stronger the magnetic fields. It is appreciated for cosmological large-scale magnetic fields. But we also see that for a single mode $k$, $a^2 B \sim \frac{H}{k^2} \sim const$ (without considering oscillation), from which it follows that the magnetic field always decreases as $\frac{1}{a^2}$ (or the energy density of the magnetic field $\rho_B \sim B^2 \propto a^{-4}$). This is the usual "adiabatic" (conformal invariance) result. At the time $\eta_{RH}$ we have $B_{RH} = \frac{e}{k_{phys}}$. It may imply that the breaking of the conformal invariance due to the minimal coupling of a massless charged scalar complex field to both gravitational and the electromagnetic fields is not sufficient for the production of seed galactic magnetic field during inflation. But since we are interested in the large-scale cosmological magnetic field, this could be still a candidate, because of the $1/k$ factor.

## 2 Preliminaries

Our current understanding of the evolution of the universe is based upon the Friedmann-Robertson-Walker (FRW) cosmological model, usually known as the hot big bang model.

### 2.1 The metric

The metric for a space with homogeneous and isotropic spatial sections is the maximally-symmetric Robert-Walker (RW) metric, which can be written in the form

$$ds^2 = g_{\mu\nu}dx^\mu dx^\nu = dx_\mu dx^\nu = dt^2 - a^2(t)[\frac{dr^2}{1-kr^2} + r^2(d^2\theta + \sin^2\theta d\varphi^2)]. \tag{2.1}$$

We use the natural units here, where $\hbar = c = k_B = 1$. Where $k = +1, -1,$ or $0$ for a closed, open, or flat (open) Friedmann universe (positive, negative or zero spatial curvature). $a(t)$ is the cosmic scale factor ("radius" of the universe), has dimensions of length.

In this paper we consider only a spatially flat universe. This refers to $k = 0$, the metric can then be put in the form

$$ds^2 = g_{\mu\nu}dx^\mu dx^\nu = dt^2 - a^2(t)\big((dx^1)^2 + (dx^2)^2 + (dx^3)^2\big). \tag{2.2}$$

At any given moment, the spatial part of the metric describes an ordinary three-dimensional Euclidean (flat) space. When $a(t)$ is constant then the flat universe metric describes Minkowski space.

### 2.2 The Friedmann's equation

Einstein's theory of gravity tell us:

$$G_{\mu\nu} = 8\pi G T_{\mu\nu} + \lambda g_{\mu\nu}. \tag{2.3}$$

Where

$$G_{\mu\nu} \equiv R_{\mu\nu} - \frac{1}{2}Rg_{\mu\nu} \tag{2.4}$$

is the Einstein Tensor and the Ricci scalar is

$$R \equiv g^{\mu\nu}R_{\mu\nu} = R^\mu_\mu. \tag{2.5}$$

And the gravitational constant $G = M_p^{-2}$, $M_p \approx 1.2 \cdot 10^{19}\,\text{GeV}$ is the Planck mass. $T_{\mu\nu}$ is the stress tensor. The simplest realization of a stress-energy tensor is characterized by a time-dependent energy density $\rho(t)$ and pressure $p(t)$ [3]:

$$T^{\mu}{}_{\nu} = \begin{pmatrix} \rho & & & \\ & -p & & \\ & & -p & \\ & & & -p \end{pmatrix}. \tag{2.6}$$

The dynamics of the expanding universe only appeared implicitly in the time dependence of the scale factor $a(t)$. To make the time dependence explicit, one must solve for the evolution of the scale factor using the Einstein's Equations (2.3).

The 0-0 component of the Einstein equation $G_{\mu\nu} = 8\pi G T_{\mu\nu}$ gives Friedmann's equation

$$\frac{\dot{a}^2 + k}{a^2} = \frac{8\pi G}{3}\rho, \tag{2.7}$$

while the $i-i$ component gives

$$2\frac{\ddot{a}}{a} + \frac{\dot{a}^2 + k}{a^2} = -8\pi G p. \tag{2.8}$$

These equations were derived in 1922 by the Russian physicist and mathematician Alexandr Friedmann (1888 - 1925), seven years before Hubble's discovery [2].

Because $\nabla_\mu T^{\mu\nu} = 0$ the (2.8) can take a easier form [7]:

$$d\rho = -(p+\rho)\frac{3da}{a}. \tag{2.9}$$

In 1917 Einstein introduced the constant Lorentz-invariant term $\lambda g_{\mu\nu}$ into his law of gravitation (2.3). This cosmological constant $\lambda$ corresponds to a tiny but universal force acting on matter. With this addition, Friedmann's equation takes the form

$$\frac{\dot{a}^2 + k}{a^2} = \frac{8\pi G}{3}\rho + \frac{\lambda}{3}, \tag{2.10}$$

and the $i-i$ component gives

$$2\frac{\ddot{a}}{a} + \frac{\dot{a}^2 + k}{a^2} = -8\pi G p + \lambda. \tag{2.11}$$

## 2.3  De Sitter space

Before going on to the general case, it is worthwhile looking at a special case for which Einstein's equation can be solved. Consider a homogeneous spatially flat universe with the Robertson-Walker metric. Friedmann's equation for the rate of expansion including the cosmological constant then takes the form

$$\frac{\dot{a}(t)}{a(t)} = H \,. \tag{2.12}$$

Where $H$ is now

$$H = \sqrt{\frac{8\pi G}{3}\rho + \frac{\lambda}{3}} \,. \tag{2.13}$$

Obviously the solution of the equation (2.12) is

$$a(t) \propto e^{Ht} \,. \tag{2.14}$$

This is an exponentially expanding universe.

In 1917 Willem de Sitter (1872 - 1934) published such a solution, setting

$$\rho = p = 0 \,. \tag{2.15}$$

Thus relating $H$ directly to the cosmological constant $\lambda$

$$H = \sqrt{\frac{\lambda}{3}} \,. \tag{2.16}$$

Eddington characterized the de Sitter universe as "motion without matter".

De Sitter space is most easily represented as a hyperboloid

$$z_0^2 - z_1^2 - z_2^2 - z_3^2 - z_4^2 = -H^{-2} \tag{2.17}$$

in five dimensional Minkowski space $(z_0, z_1, z_2, z_3, z_4)$. To facilitate an intuitive interpretation of a curved four-dimensional space, it is often convenient to imagine it as a curved four-dimensional hypersurface embedded in a higher-dimensional space.

In order to represent de Sitter space as a flat Friedmann universe, it suffices to consider a coordinate system $t, x_i$ on the hyperboloid (2.17), defined by [1]

$$\begin{aligned} z_0 &= H^{-1} \sinh Ht + \frac{1}{2} He^{Ht} \vec{x}^2 \\ z_4 &= H^{-1} \cosh Ht - \frac{1}{2} He^{Ht} \vec{x}^2 \\ z_i &= e^{Ht} x_i, \qquad i = 1, 2, 3 \end{aligned} \tag{2.18}$$

The branch corresponding to that Robertson-Walker space-time spans the half hyperboloid by

$$z_0 + z_4 > 0 \,. \tag{2.19}$$

And the metric takes the form

$$ds^2 = dt^2 - e^{2Ht} d\vec{x}^2 = dt^2 - a^2(t) d\vec{x}^2 \,. \tag{2.20}$$

## 3    The generation of the cosmological magnetic fields

It has been suggested [9] that the magnetic fields observed in the galaxy and in extragalactic objects may have existed before galaxies formed, perhaps originating in primeval turbulence at high red shift, perhaps present at the time of the big bang. It has also been argued that the magnetic fields of the Milky Way galaxy could not persist for ~10 G yr because it is dynamically unstable against leaving the galaxy in a shorter time. If so, a dynamo is needed to maintain the field in the galaxy and maybe able to generate the field in the first place from a very weak seed (Parker 1975) [9].

An interesting consequence of weak tangled primeval field is discussed by Wasserman (1978): the field acts as a source of density irregularities that grow through ordinary gravitational instability. Thus if a tangled primeval field is postulated, it implies a minimum value of $\delta\rho/\rho$ (p.71 [9]).

Many astrophysicists believe that galactic magnet fields are generated and maintained by dynamo action (where by the energy associated with the differential rotation of spiral galaxies is converted into magnetic field energy) [5]. The dynamo mechanism is only a means of amplification and dynamos require seed magnetic fields. If a galactic dynamo has operated over the entire age of the galaxy $(\sim 10 \text{ G yr})$, it could have amplified a seed field by a factor of $e^{[O(30)]}$, implying that a seed magnetic field $\sim 3 \times 10^{-19}$ G is required [5]. Some astrophysicists believe the galactic magnetic field owes its existence to primeval magnetic flux trapped in the gas that collapsed to form the galaxy [5]; in this case the primeval fields strength required is much greater ─ at least the field strength that is observed today $\sim 3 \times 10^{-6}$ G.

Harrison has proposed a mechanism for producing the small seed field required for the galactic dynamo, wherein the relative motions of protons and electrons induced by vorticity present prior to decoupling produce primeval currents and magnetic fields ─ of course, this presupposes the existence of primeval vorticity. Other scenarios have also been suggested [5]. A fair summary of the present situation is to say that no compelling mechanism has yet been suggested for the origin of the essential primeval magnetic fields [5]. Turner and Widrow believe that inflation is a prime candidate for the production of primeval magnetic fields [5].

## 4 The Maxwell equation and the wave equation

Before discussing the evolution of the cosmological magnetic fields we review the Maxwell equation in curved space-time. After that we develop it in a homogeneous spatially flat universe with the Robertson-Walker metric. Based on this we develop the wave equation of the cosmological magnetic fields. This wave equation gives us a basis for the further studying of the evolution of the cosmological magnetic fields in Friedmann-Robertson-Walker spatially flat universe.

### 4.1 The Maxwell equation

We are going to work out the Maxwell equation in a homogeneous spatially flat universe with the Robertson-Walker metric. Let us start with the general form of the Maxwell equation in curved space-time.

#### 4.1.1 The Maxwell equation in curved space-time

We begin with the Maxwell equation

$$F_{\mu\nu;\sigma} + F_{\sigma\mu;\nu} + F_{\nu\sigma;\mu} = 0$$
$$F^{\mu\nu}{}_{;\mu} = J^{\nu} \qquad (4.1)$$

One regards the field tensor $F_{\mu\nu}$ as arising from a potential by the

$$F_{\mu\nu} = A_{\nu;\mu} - A_{\mu;\nu} . \qquad (4.2)$$

These points granted, one can verify that second of Maxwell's equations is automatically satisfied and verify also that the first is satisfied if and only if

$$-A^{\nu;\mu}{}_{\mu} + A^{\mu}{}_{;\mu}{}^{\nu} + R^{\nu}{}_{\mu}A^{\mu} = J^{\nu} . \qquad (4.3)$$

Where $A^{\mu}$ is the 4-components potential and $J^{\mu}$ the 4-components current.

We recall in a flat space-time, one can always find a coordinate system so that the connection coefficients $\Gamma^{\lambda}{}_{\mu\nu} = 0$. The covariant derivatives then becomes to normal derivatives

$$F_{\mu\nu,\sigma} + F_{\sigma\mu,\nu} + F_{\nu\sigma,\mu} = 0$$
$$F^{\mu\nu}{}_{,\mu} = J^{\nu} \qquad (4.4)$$

Where

$$F_{\mu\nu} = A_{\nu,\mu} - A_{\mu,\nu} = \begin{pmatrix} 0 & E_1 & E_2 & E_3 \\ -E_1 & 0 & -B_3 & B_2 \\ -E_2 & B_3 & 0 & -B_1 \\ -E_3 & -B_2 & B_1 & 0 \end{pmatrix} = -F_{\nu\mu} \qquad (4.5)$$

is anti-symmetric in Minkowski space.

In a general way, curved space-time, we may rewrite the Maxwell equation as follows

$$\begin{aligned}&\frac{\partial}{\partial x^\sigma}F_{\mu\nu}+\frac{\partial}{\partial x^\nu}F_{\sigma\mu}+\frac{\partial}{\partial x^\mu}F_{\nu\sigma}=0\\&\frac{\partial}{\partial x^\mu}\sqrt{-g}F^{\mu\nu}=\sqrt{-g}J^\nu\end{aligned}. \tag{4.6}$$

Here we notice, we use the fact that $F_{\mu\nu}$ and $F^{\mu\nu}$ are antisymmetric and so they obey the following equations [12]

$$T^{\mu\nu}{}_{;\mu}=\frac{1}{\sqrt{-g}}\frac{\partial}{\partial x^\mu}\left(\sqrt{-g}T^{\mu\nu}\right) \qquad \text{for } T^{\mu\nu} \text{ antisymmetric} \tag{4.7}$$

$$T_{\mu\nu;\sigma}+T_{\sigma\mu;\nu}+T_{\nu\sigma;\mu}=\frac{\partial}{\partial x^\sigma}T_{\mu\nu}+\frac{\partial}{\partial x^\nu}T_{\sigma\mu}+\frac{\partial}{\partial x^\mu}T_{\nu\sigma} \qquad \text{for } T^{\mu\nu} \text{ antisymmetric}. \tag{4.8}$$

### 4.1.2   The Maxwell equation in FRW spatially flat universe

Now in our Friedmann-Robertson-Walker spatially flat universe we recall here formula (2.2) $ds^2 = dt^2 - a^2(t)\left((dx^1)^2 + (dx^2)^2 + (dx^3)^2\right)$. With conformal time $\eta$ (light cones at $45°$) we have

$$ds^2 = dt^2 - a^2(t)d\vec{x}^2 \equiv a^2(\eta)(d\eta^2 - d\vec{x}^2). \tag{4.9}$$

Where

$$dt = ad\eta \;\Rightarrow\; \eta = \int \frac{dt}{a(t)}. \tag{4.10}$$

Here the metric $g_{\mu\nu}$ is

$$g_{\mu\nu} = a^2(\eta)\begin{pmatrix} 1 & & & \\ & -1 & & \\ & & -1 & \\ & & & -1 \end{pmatrix}. \tag{4.11}$$

Then we have the field tensor

$$F_{\mu\nu} = a^2(\eta)\begin{pmatrix} 0 & E_1 & E_2 & E_3 \\ -E_1 & 0 & -B_3 & B_2 \\ -E_2 & B_3 & 0 & -B_1 \\ -E_3 & -B_2 & B_1 & 0 \end{pmatrix} = -F_{\nu\mu}. \tag{4.12}$$

We use the equation (4.10) $dt = ad\eta$.

We define now

$$g \equiv \det g_{\mu\nu}. \tag{4.13}$$

In our case we see

$$g = -a^8(\eta). \tag{4.14}$$

We know

$$F_{\mu\nu} = g_{\mu\lambda} g_{\nu\kappa} F^{\lambda\kappa} \tag{4.15}$$

and get

$$F^{\mu\nu} = \frac{1}{a^2(\eta)} \begin{pmatrix} 0 & -E_1 & -E_2 & -E_3 \\ E_1 & 0 & -B_3 & B_2 \\ E_2 & B_3 & 0 & -B_1 \\ E_3 & -B_2 & B_1 & 0 \end{pmatrix} = -F^{\nu\mu}. \tag{4.16}$$

Before we write down the Maxwell equation (4.6) in components, let us see the relationships between the tensor analysis formalism outlined above and the familiar formulas for gradient, curl, and divergence in the classical curvilinear coordinate systems at first. Details are in the Appendix A, and we write down the result for our FRW spatially flat universe:

$$\text{div}\, \vec{A} = \frac{1}{a}\left(\frac{\partial A_{\hat{1}}}{\partial x^1} + \frac{\partial A_{\hat{2}}}{\partial x^2} + \frac{\partial A_{\hat{3}}}{\partial x^3}\right), \tag{4.17}$$

$$(\text{curl}\,\vec{A})_1 = \frac{1}{a}\left(\frac{\partial A_{\hat{3}}}{\partial x^2} - \frac{\partial A_{\hat{2}}}{\partial x^3}\right) \quad \text{(and cyclical)}. \tag{4.18}$$

We notice here, $A_{\hat{i}}$ are the "ordinary" components $A_{\hat{i}} \equiv \vec{A} \cdot \vec{e}_{\hat{i}}$ (means physically measured in LONB). Without other declaration, the components like $E_i, B_i$ in this paper means the "ordinary" components $E_{\hat{i}}, B_{\hat{i}}$.

So we have now the Maxwell equations as follows

$$\begin{aligned} \frac{\partial}{\partial t} a^2 \vec{B} + a^2\, \text{curl}\, \vec{E} &= 0 \\ \text{div}\, \vec{B} &= 0 \end{aligned} \tag{4.19}$$

and

$$\begin{aligned} \text{curl}\, \vec{B} - \frac{1}{a^2}\frac{\partial}{\partial t} a^2 \vec{E} &= \vec{j} \\ \text{div}\, \vec{E} &= \rho \end{aligned} \tag{4.20}$$

For the conformal time $\eta$, we have

$$\frac{1}{a^3}\frac{\partial}{\partial \eta}a^2\vec{B} + \operatorname{curl}\vec{E} = 0$$
$$\operatorname{div}\vec{B} = 0 \tag{4.21}$$

and

$$\operatorname{curl}\vec{B} - \frac{1}{a^3}\frac{\partial}{\partial \eta}a^2\vec{E} = \vec{j}$$
$$\operatorname{div}\vec{E} = \rho \tag{4.22}$$

We notice here the $\rho, \vec{j}$ are "ordinary" and the definition of $\eta, t$ are as follows

$$\eta \equiv x^0$$
$$dt \equiv a d\eta \tag{4.23}$$

$$\rho \equiv aJ^0$$
$$\vec{j} \equiv a\left(J^1, J^2, J^3\right) \tag{4.24}$$

### 4.2 The wave equation of the magnetic fields

From (4.20) we can do more for the magnetic fields

$$\operatorname{curl}\vec{B} = \frac{1}{a^2}\frac{\partial}{\partial t}a^2\vec{E} + \vec{j}. \tag{4.25}$$

We then use the formula $\operatorname{curl}\operatorname{curl}\vec{v} = \operatorname{grad}(\operatorname{div}\vec{v}) - \Delta\vec{v}$ and we know $\operatorname{div}\vec{B} = 0$. Here we get

$$-\Delta\vec{B} = \operatorname{curl}\left(\frac{1}{a^2}\frac{\partial}{\partial t}a^2\vec{E} + \vec{j}\right)$$
$$= \frac{1}{a^2}\operatorname{curl}\left(\frac{\partial a^2\vec{E}}{\partial t}\right) + \operatorname{curl}\vec{j} \tag{4.26}$$

We define here

$$\dot{a} \equiv \frac{\partial a}{\partial t}. \tag{4.27}$$

Before we go further, we want to calculate the commutation relations of $\operatorname{curl}$ and $\frac{\partial}{\partial t}$ for an arbitrary vector $\vec{A}$

$$\left(\left[\mathrm{curl}, \frac{\partial}{\partial t}\right]\vec{A}\right)_1 = \left(\mathrm{curl}\left(\frac{\partial}{\partial t}\vec{A}\right) - \frac{\partial}{\partial t}\left(\mathrm{curl}\,\vec{A}\right)\right)_1$$
$$= \frac{1}{a}(\frac{\partial \dot{\overline{A}}_3}{\partial x^2} - \frac{\partial \dot{\overline{A}}_2}{\partial x^3}) - \frac{\partial}{\partial t}\left(\frac{1}{a}(\frac{\partial \overline{A}_3}{\partial x^2} - \frac{\partial \overline{A}_2}{\partial x^3})\right)$$
$$= \frac{\dot{a}}{a^2}\left[(\frac{\partial \overline{A}_3}{\partial x^2} - \frac{\partial \overline{A}_2}{\partial x^3})\right]$$
$$= \frac{\dot{a}}{a}\left(\mathrm{curl}\,\vec{A}\right)_1 \qquad (4.28)$$

So we get

$$-\Delta \vec{B} = \frac{1}{a^2}\left(\frac{\partial}{\partial t}\left(\mathrm{curl}\,a^2\vec{E}\right) + \frac{\dot{a}}{a}\mathrm{curl}\,a^2\vec{E}\right) + \mathrm{curl}\,\vec{j}\,. \qquad (4.29)$$

From (4.19) we know:

$$\mathrm{curl}\,a^2\vec{E} = a^2\,\mathrm{curl}\,\vec{E} = -\frac{\partial}{\partial t}(a^2\vec{B})\,. \qquad (4.30)$$

Now we put the equation (4.30) to (4.29)

$$\Delta \vec{B} = -\left(\frac{1}{a^2}\left(\frac{\partial}{\partial t}\left(-\frac{\partial}{\partial t}(a^2\vec{B})\right) + \frac{\dot{a}}{a}\left(-\frac{\partial}{\partial t}(a^2\vec{B})\right)\right) + \mathrm{curl}\,\vec{j}\right)$$
$$= \frac{1}{a^2}\frac{\partial^2 a^2\vec{B}}{\partial t^2} + \frac{\dot{a}}{a^3}\frac{\partial a^2\vec{B}}{\partial t} - \mathrm{curl}\,\vec{j} \qquad (4.31)$$

and we know

$$H \equiv \frac{\partial a}{\partial t}\Big/a = \frac{\dot{a}}{a}$$
$$= \frac{\partial a}{a\partial \eta}\Big/a = \frac{1}{a^2}\frac{\partial a}{\partial \eta}\,. \qquad (4.32)$$

So now we put this into equation (4.31) and get

$$\Delta \vec{B} - \frac{1}{a^2}\frac{\partial a^2 \vec{B}}{\partial t^2} - \frac{H}{a^2}\frac{\partial a^2 \vec{B}}{\partial t} + \mathrm{curl}\,\vec{j} = 0\,. \qquad (4.33)$$

And this is the wave equation of magnetic Field.

With conformal time $\eta$,

$$\frac{\partial^2}{\partial t^2} = \frac{1}{a}\frac{\partial}{\partial \eta}\left(\frac{1}{a}\frac{\partial}{\partial \eta}\right)$$
$$= \frac{1}{a^2}\frac{\partial^2}{\partial \eta^2} - \frac{1}{a^3}\frac{\partial a}{\partial \eta}\frac{\partial}{\partial \eta} \quad . \tag{4.34}$$
$$= \frac{1}{a^2}\frac{\partial^2}{\partial \eta^2} - \frac{H}{a}\frac{\partial}{\partial \eta}$$

On the other hand

$$\frac{\partial^2}{\partial \eta^2} = a\frac{\partial}{\partial t}\left(a\frac{\partial}{\partial t}\right)$$
$$= a^2 \frac{\partial^2}{\partial t^2} + a\dot{a}\frac{\partial}{\partial t} \quad . \tag{4.35}$$
$$= a^2 \frac{\partial^2}{\partial t^2} + Ha^2\frac{\partial}{\partial t}$$

And the wave equation (4.33) can be written as follows

$$\Delta \vec{B} - \frac{1}{a^4}\frac{\partial^2\left(a^2\vec{B}\right)}{\partial \eta^2} + \operatorname{curl}\vec{j} = 0 \,. \tag{4.36}$$

We see if we know the initial conditions of $\vec{B}$ (spatial distribution of $\vec{B}(0)$ at initial time and $\dot{\vec{B}}(0)$), $a$ and the current $\vec{j}$, we can then calculate the evolution of the magnetic field.

Similarly we get another equation for $\vec{E}$ (Details in Appendix B)

$$\Delta \vec{E} - \frac{1}{a^2}\frac{\partial^2\left(a^2\vec{E}\right)}{\partial t^2} - \frac{H}{a^2}\frac{\partial\left(a^2\vec{E}\right)}{\partial t} - \frac{1}{a^2}\frac{\partial\left(a^2\vec{j}\right)}{\partial t} - H\vec{j} - \operatorname{grad}\rho = 0 \,, \tag{4.37}$$

with conformal time $\eta$, we have

$$\Delta \vec{E} - \frac{1}{a^4}\frac{\partial^2\left(a^2\vec{E}\right)}{\partial \eta^2} - \frac{1}{a^3}\frac{\partial\left(a^2\vec{j}\right)}{\partial \eta} - H\vec{j} - \operatorname{grad}\rho = 0 \,. \tag{4.38}$$

## 5 The correlation function and the current

In the early universe gravity acting on a quantum matter field amplifies energy density fluctuations above the unavoidable vacuum fluctuations. These primordial fluctuations may serve as seeds for the subsequent formation of the observable large-scale structure [16].

### 5.1 The Model

Let us consider a charged complex scalar field $\phi$ and the lagrangian density will be

$$\mathscr{L} = \sqrt{-g}\left(D_\mu\phi(D^\mu\phi)^* - \frac{1}{4}F_{\mu\nu}F^{\mu\nu} - [m^2 + \xi R]\phi^2\right). \tag{5.1}$$

The coupling between the scalar field and the gravitational field is represented by the term $\xi R\phi^2$, where $\xi$ is a numerical factor and $R$ is the Ricci scalar. For simplicity we are neglecting the field's coupling to other fields. The complex scalar field couples to electromagnetism through the usual gauge covariant derivative

$$D_\mu = \partial_\mu - ieA_\mu. \tag{5.2}$$

The current is defined as following

$$J_\mu = ie\left(\phi^*\overleftrightarrow{D}_\mu\phi\right) = ie\left(\phi^*D_\mu\phi - \phi D_\mu\phi^*\right). \tag{5.3}$$

And the conservation equation for the current is given by [13]

$$\frac{1}{\sqrt{-g}}\partial_\mu(\sqrt{-g}J^\mu) = 0. \tag{5.4}$$

The Klein-Gordon equation in curved space-time is [11]

$$\left(\Box + m^2 + \xi R\right)\phi = 0. \tag{5.5}$$

In the minimally coupled case,

$$\xi = 0. \tag{5.6}$$

The action $S = \int d^4x\mathscr{L}$ of a charged scalar field minimally coupled to the background geometry and to the electromagnetic field is then [13]

$$S = \int d^4x\sqrt{-g}\left((D_\mu\phi)^*D^\mu\phi - m\phi^*\phi - \frac{1}{4}F_{\alpha\beta}F^{\alpha\beta}\right). \tag{5.7}$$

Where $D_\mu = \partial_\mu - ieA_\mu$, $F_{\mu\nu} = A_{\nu,\mu} - A_{\mu,\nu}$.

In the case $A_\mu = 0$ and $m = 0$, the action is simply given by

$$S = \int d^4x \sqrt{-g}\left((\partial_\mu \phi)^* \partial^\mu \phi\right). \tag{5.8}$$

We consider here in de Sitter space-time:

$$ds^2 = dt^2 - e^{2Ht}d\vec{x}^2 = dt^2 - a^2(t)d\vec{x}^2. \tag{5.9}$$

With conformal time $\eta$ (light cones at $45°$) we have

$$ds^2 = dt^2 - a^2(t)d\vec{x}^2 \equiv a^2(\eta)(d\eta^2 - d\vec{x}^2). \tag{5.10}$$

Where

$$dt = ad\eta \Rightarrow \eta = \int \frac{dt}{a(t)}. \tag{5.11}$$

The energy density $\rho$ is always defined to be the one measured by a commoving observer, who has the 4-velocity $u^\mu = (a^{-1}, 0, 0, 0)$ in conformal coordinates [16]. Thus we obtain [16]

$$\rho = \frac{1}{2a^2}\left[(\partial_\eta \phi)^2 + (\partial_{\vec{x}} \phi)^2 + a^2 m^2 \phi^2\right]. \tag{5.12}$$

For $m = 0$ we get

$$\rho = \frac{1}{2a^2}\left[(\partial_\eta \phi)^2 + (\partial_{\vec{x}} \phi)^2\right]. \tag{5.13}$$

The fluctuations in the energy density $\rho$ are given by the two-point correlation function at equal time:

$$\begin{aligned}C(l) \equiv C(\vec{x}, \vec{x}') &= \langle \rho(\vec{x})\rho(\vec{x}') \rangle - \langle \rho(\vec{x}) \rangle \langle \rho(\vec{x}') \rangle \\ &= \langle \psi | \rho(\vec{x})\rho(\vec{x}') | \psi \rangle - \langle \psi | \rho(\vec{x}) | \psi \rangle \langle \psi | \rho(\vec{x}') | \psi \rangle\end{aligned}. \tag{5.14}$$

Where $l$ is the measured distance between the two points $\vec{x}$ and $\vec{x}'$:

$$\vec{l} \equiv \vec{x}_{phys} - \vec{x}'_{phys} = a(\vec{x} - \vec{x}') \tag{5.15}$$

and $l \equiv |\vec{l}|$.

During inflation the Hubble parameter is constant, and one has de Sitter space-time. The end of the de Sitter era is denoted by RH for reheating. With $a(t) = a(t_{RH})e^{H(t-t_{RH})}$, we can then fix the inflationary scale factor by setting the followings at time $t_{RH}$ [15]:

$$a(t_{RH}) \equiv 1, \eta_{RH} \equiv -H^{-1}. \tag{5.16}$$

So we get from (5.11)

$$\eta(t_{RH}) - \eta(t) = \int \frac{dt}{a(t)} = \int_t^{t_{RH}} dt' e^{-H(t'-t_{RH})} = -\frac{1}{H}\left(1 - e^{-H(t-t_{RH})}\right). \tag{5.17}$$

We now get

$$\eta(t) = \eta(t_{RH}) - \left(-\frac{1}{H}\left(1 - e^{-H(t-t_{RH})}\right)\right) = \left(-\frac{1}{H}\right) + \frac{1}{H} - \frac{e^{-H(t-t_{RH})}}{H}$$
$$= -\frac{e^{-H(t-t_{RH})}}{H} \tag{5.18}$$

From (5.11) we see

$$dt = a(\eta)d\eta \Rightarrow a(\eta) = \frac{dt}{d\eta}$$
$$\Rightarrow a(\eta) = 1 \Big/ \frac{d\eta}{dt} \tag{5.19}$$

With (5.18) we see

$$\frac{d\eta}{dt} = (-H)\left(-\frac{e^{-H(t-t_{RH})}}{H}\right).$$
$$= -H\eta \tag{5.20}$$

So we get

$$a(\eta) = 1\Big/\frac{d\eta}{dt} = -\frac{1}{H\eta}. \tag{5.21}$$

The Klein-Gordon equation is from the Variation of the action $S$, this is [8]:

$$\left[\partial_\eta^2 + 2\left(\frac{\partial_\eta a}{a}\right)\partial_\eta - \partial_i^2 + m^2 a^2\right]\phi = 0. \tag{5.22}$$

The mode functions in a spatially flat FRW space-time are eigenfunctions of the commoving wave vector $\vec{k}$. Because of the translational invariance of the gravitational field in 3-space different $\vec{k}$'s decouple, and $\vec{k}$ is a conserved quantity [16]. We make the ansatz of modes of fixed $\vec{k}$

$$\varphi_k(\eta, \vec{x}) \equiv u_k(\eta)e^{i\vec{k}\vec{x}}. \tag{5.23}$$

We notice here:

$$\vec{k} \equiv (k_1, k_2, k_3), \quad \vec{k}_{phys} = \vec{k}/a$$
$$\omega \equiv k_0, \qquad \omega_{phys} = \omega/a. \tag{5.24}$$
$$\vec{x} \equiv (x^1, x^2, x^3)$$

We put (5.23) into (5.22) using the equation (5.21) and obtain the equation of motion for the different $u_k$

$$\left[\partial_\eta^2 - \frac{2}{\eta}\partial_\eta + k^2 + \frac{m^2}{H^2\eta^2}\right]u_k(\eta) = 0. \tag{5.25}$$

Its solutions are linear combinations of Hankel functions $h_\nu^{(1,2)}$ times $\eta^{3/2}$ [15],

$$u_k(\eta) = const.\cdot \eta^{3/2}\cdot h_\nu^{(1,2)}(k\eta), \quad \nu = \sqrt{\frac{9}{4} - \frac{m^2}{h^2}}. \tag{5.26}$$

For the basis modes $\varphi_k(\eta,\vec{x})$ we choose that fundamental solution in (5.26) which has the time dependence $e^{-ik\eta}$ for $\eta \to -\infty$, i.e. $h_\nu^{(2)}(k\eta)$. The solution approaches a Minkowski single-particle wave at early time, when the physical wavelength is much smaller than the Hubble radius. Therefore one could write more explicitly $\varphi_k(\eta,\vec{x}) = \varphi_k^{(in)}(\eta,\vec{x})$. [15]

The normalization of the mode functions in (5.23) is as follows

$$(\varphi_k, \varphi_{k'}) = (2\pi)^3 \delta^{(3)}(\vec{k} - \vec{k}'). \tag{5.27}$$

With respect to the scalar product [15]

$$(\varphi_k, \varphi_{k'}) \equiv i\int d^3x \sqrt{-g^{(3)}}(\varphi_k^* \overleftrightarrow{\partial_{\hat 0}} \varphi_{k'}). \tag{5.28}$$

Where $\partial_{\hat 0}$ denotes the derivative with respect to measured time along the normal to the slice $t = const.$, and the 3-dimensional $\sqrt{-g^{(3)}}$ takes the value $\sqrt{-g^{(3)}} = a^3$.

The normalized basis modes are given by [15]

$$\begin{aligned}\varphi_k(\eta,\vec{x}) &= \frac{H}{\sqrt{2k^3}}(k\eta - i)e^{i(\vec{k}\vec{x} - k\eta)}, \quad \text{for } m = 0 \\ \varphi_k(\eta,\vec{x}) &= \frac{\sqrt{\pi}H}{2}\eta^{3/2}h_\nu^{(2)}(k\eta)e^{i\vec{k}\vec{x}}, \quad \text{for } m \neq 0\end{aligned}. \tag{5.29}$$

The mode expansion of the charged complex scalar field operator $\phi(\eta,\vec{x})$ is

$$\phi(\eta,\vec{x}) = \int \frac{d^3k}{(2\pi)^3}\left(\varphi_k(\eta,\vec{x})a_k + \varphi_k^*(\eta,\vec{x})b_k^\dagger\right). \tag{5.30}$$

By convention, we refer to an **a** quantum as a "particle" and a **b** quantum as an "antiparticle". Thus, the positive-frequency part of $\phi$ annihilates a particle, and its negative-frequency part creates an antiparticle. Similarly, $\phi^*$ either creates a particle or annihilates an antiparticle. The canonical commutation relations $[\phi(\vec{x}), \Pi(\vec{y})] = i\delta^{(3)}(\vec{x} - \vec{y})$, where $\Pi = a^3\partial_t\phi$, fix the normalization factor in the commutation relations for the annihilation and creation operators [15],

$$\left[a_k, a_{k'}^\dagger\right] = (2\pi)^3\delta^{(3)}(\vec{k} - \vec{k}'). \tag{5.31}$$

The initial state of the quantum field is dictated by the inflationary paradigm:

1.) Every mode with comoving wavenumber $k$ which is observable today had a physical wavelength much smaller than the Hubble radius at early enough times during the inflationary era, i.e. it starts out sub-horizon-sized $\lambda_{phys} \ll H^{-1}$, the space-time curvature $R$ was much smaller than $k_{phys}^2$ and negligible for the physically relevant modes. [15]

2.) The inflationary expansion wiped out all perturbations except the unavoidable quantum fluctuations. For every relevant mode the initial state is the vacuum state (ground state) with its inescapable vacuum fluctuations. In the Heisenberg picture of time evolution the states are defined to be time independent and denoted by the properties of the initial state, in our case $|0\ in\rangle$ for all relevant modes (observable today). This state of the quantum fields is the Bunch-Davies state. [15]

The Heisenberg state $|0\ in\rangle$ is annihilated by the time-independent annihilation operators $a_k$ and $b_k$ of the mode expansion (5.30). [15]

$$\begin{aligned} a_k |0\ in\rangle &= 0 \\ b_k |0\ in\rangle &= 0 \end{aligned} \quad . \tag{5.32}$$

We also know

$$\langle 0|0\rangle = 1 . \tag{5.33}$$

We consider now on super-horizon limes $\lambda_{phys} \gg R_H = \frac{1}{H}$ it also means

$$\begin{aligned} \frac{1}{k_{phys}} &\gg \frac{1}{H} \quad \text{or} \\ k_{phys}\frac{1}{H} &\ll 1 \end{aligned} \quad . \tag{5.34}$$

We notice here with the equation (5.21) $a(\eta) = -\frac{1}{H\eta}$, one gets

$$k_{phys}\frac{1}{H} = \frac{k}{a}\frac{1}{H} = -k\eta = |k\eta| . \tag{5.35}$$

We know here is $\eta < 0$. So we get for super-horizon limes is

$$|k\eta| \ll 1 \tag{5.36}$$

and for sub-horizon limes is

$$|k\eta| \gg 1 . \tag{5.37}$$

For the crossing area we define now the $\eta_{HC}$ as follows

$$(k\eta)_{HC} \equiv -1. \tag{5.38}$$

## 5.2 The current

Before we start in de Sitter space-time let us see the situation in Minkowski space-time

$$\begin{aligned}\phi(t,\vec{x}) &= \int \frac{d^3k}{(2\pi)^3} \Big( e^{i(\vec{k}\vec{x}-\omega t)} a_{k,in} + e^{-i(\vec{k}\vec{x}-\omega t)} b^\dagger_{k,in} \Big) \\ &= \int \frac{d^3k}{(2\pi)^3} \Big( e^{-ik_\mu x^\mu} a_{k,in} + e^{ik_\mu x^\mu} b^\dagger_{k,in} \Big)\end{aligned} \tag{5.39}$$

Where

$$\begin{aligned}k_\mu &= (\omega, -\vec{k}) \\ x^\mu &= (t, \vec{x})\end{aligned} \tag{5.40}$$

In the case $A^\mu = 0$

$$J_\mu = ie : \phi^* \overleftrightarrow{\partial}_\mu \phi := ie\big(\phi^* \partial_\mu \phi - \phi \partial_\mu \phi^*\big). \tag{5.41}$$

This is the Noether current for the $U(1)$ symmetry of the action (5.8). And we obtain

$$\begin{aligned}&\left\langle \pi^+_{\vec{k}'} \big| J_\mu(\vec{x},t) \big| \pi^+_{\vec{k}} \right\rangle \\ &= \int \frac{d^3q}{(2\pi)^3} \int \frac{d^3q'}{(2\pi)^3} \left\langle \pi^+_{\vec{k}'} \big| a^\dagger_{q'} a_q \big| \pi^+_{\vec{k}} \right\rangle \cdot e(q_\mu + q'_\mu) e^{-i(q_\mu - q'_\mu)x^\mu} \\ &= e(k_\mu + k'_\mu) e^{-i(k_\mu - k'_\mu)x^\mu}\end{aligned} \tag{5.42}$$

Where we use the commutation relations (5.31) to get the following relation

$$\begin{aligned}a_{\vec{k}} \ket{\vec{p}} &= a_{\vec{k}} a^\dagger_{\vec{p}} \ket{0} = \big(a^\dagger_{\vec{p}} a_{\vec{k}} + (2\pi)^3 \delta^{(3)}(\vec{k}-\vec{p})\big) \ket{0} \\ &= (2\pi)^3 \delta^{(3)}(\vec{k}-\vec{p}) \ket{0}\end{aligned} \tag{5.43}$$

And we can also obtain $\mathrm{curl}\,\vec{j}$ from (5.42)

$$\begin{aligned}\left\langle \pi^+_{\vec{k}'} \big| \mathrm{curl}\,\vec{j} \big| \pi^+_{\vec{k}} \right\rangle &= \left\langle \pi^+_{\vec{k}'} \big| \nabla \times \vec{j} \big| \pi^+_{\vec{k}} \right\rangle \\ &= ie(\vec{k}-\vec{k}') \times (\vec{k}+\vec{k}') e^{-i(k_\mu - k'_\mu)x^\mu}\end{aligned} \tag{5.44}$$

We can also write it in component

$$\begin{aligned}&\left\langle \pi^+_{\vec{k}'}, in \big| (\mathrm{curl}\,\vec{j})_1 \big| \pi^+_{\vec{k}}, in \right\rangle \\ &= 2ie(k_2 k'_3 - k'_2 k_3) e^{-i(k_\mu - k'_\mu)x^\mu} \quad \text{and cyclical}\end{aligned} \tag{5.45}$$

Now in de Sitter space-time we recall (4.17) and (4.18) for the calculation rules and we get

$$\left\langle \pi^+_{\vec{k}_+}, \pi^-_{\vec{k}_-} \middle| \vec{j} \middle| 0 \; in \right\rangle$$
$$= \frac{1}{a} \int \frac{d^3q}{(2\pi)^3} \int \frac{d^3q'}{(2\pi)^3} \left\langle \pi^+_{\vec{k}_+}, \pi^-_{\vec{k}_-} \middle| a^\dagger_q b^\dagger_{q'} \middle| 0 \; in \right\rangle \cdot e \frac{H^2}{\sqrt{2q^3}\sqrt{2q'^3}} (q\eta - i)(q'\eta - i)(\vec{q}' - \vec{q}) e^{i(q_\mu + q'_\mu) x^\mu}$$
$$= e \frac{H^2}{a\sqrt{2k_+^3}\sqrt{2k_-^3}} (k_+\eta - i)(k_-\eta - i)(\vec{k}_- - \vec{k}_+) e^{i(k_{+\mu} + k_{-\mu})x^\mu}$$
(5.46)

We notice the $\rho, \vec{j}$ here are the ordinary components (physically measured in LONB)

$$\rho \equiv aJ^0 = J_0 / a$$
$$\vec{j} \equiv a(J^1, J^2, J^3) = (J_1, J_2, J_3)/a$$
(5.47)

and

$$\left\langle \pi^+_{\vec{k}_+}, \pi^-_{\vec{k}_-} \middle| \operatorname{curl} \vec{j} \middle| 0 \; in \right\rangle = \left\langle \pi^+_{\vec{k}_+}, \pi^-_{\vec{k}_-} \middle| \nabla \times \vec{j} \middle| 0 \; in \right\rangle$$
$$= ie \frac{H^2}{a^2 \sqrt{2k_+^3}\sqrt{2k_-^3}} (k_+\eta - i)(k_-\eta - i)(\vec{k}_+ + \vec{k}_-) \times (\vec{k}_+ - \vec{k}_-) e^{i(k_{+\mu} + k_{-\mu})x^\mu}$$
(5.48)

One can see it easily from (5.48), that for $\vec{k}_+ = -\vec{k}_-$ is $\left\langle \pi^+_{\vec{k}_+}, \pi^-_{\vec{k}_-} \middle| \operatorname{curl} \vec{j} \middle| 0 \; in \right\rangle = 0$. Now let we consider the quantity

$$\sqrt{\left| \left\langle \pi^+_{\vec{k}_+}, \pi^-_{\vec{k}_-} \middle| \vec{j} \middle| 0 \; in \right\rangle \right|^2}, \quad \vec{k}_+ = -\vec{k}_-.$$
(5.49)

From (5.46) we get

$$\sqrt{\left| \left\langle \pi^+_{\vec{k}_+}, \pi^-_{\vec{k}_-} \middle| \vec{j} \middle| 0 \; in \right\rangle \right|^2} = e \frac{H^2}{2ak^3} \sqrt{\left|(k\eta - i)^2\right|^2} 2k$$
$$= e \frac{H^2}{ak^2} \sqrt{(k^2\eta^2 - 1)^2 + (2k\eta)^2} = e \frac{H^2}{ak^2} (1 + k^2\eta^2).$$
$$= e \frac{H^2}{ak^2} + \frac{eH^2\eta^2}{a}$$
(5.50)

When on super-horizon limes $\lambda \gg R_H = \frac{1}{H}$, one sees directly that the first term is dominated, i.e., the current depends on $k$. The graphic are as follows:

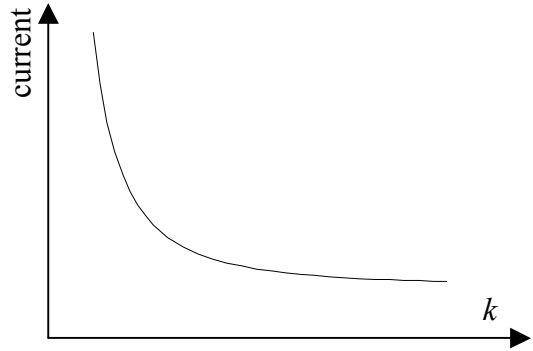

Figure (4.1) The current on super-horizon limes

With logarithmic coordinates system, we have:

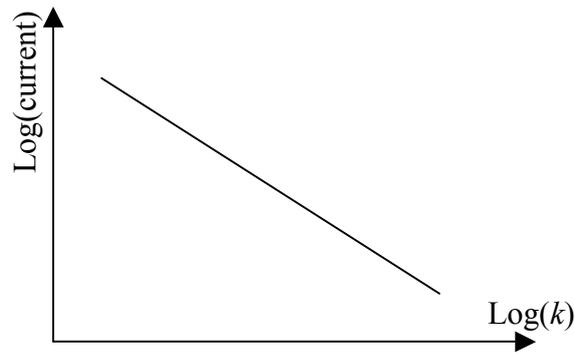

Figure (4.2) The current on super-horizon limes

For sub-horizon limes, the second term is dominated, and it's growth quadratic with time.

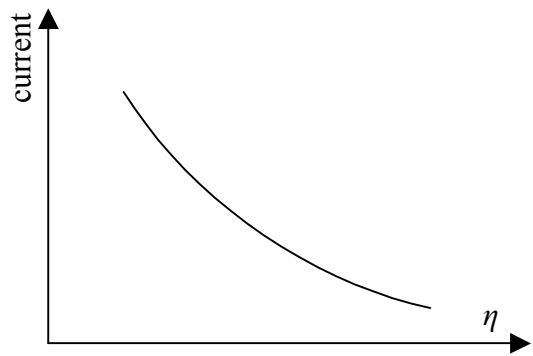

Figure(4.3) The current on sub-horizon limes

We notice here that on (5.16) we have set $\eta_{RH} \equiv -H^{-1}$ and therefore in our case $\eta < 0$.

For a fixed $k$, and with a logarithmic coordinate system, we have:

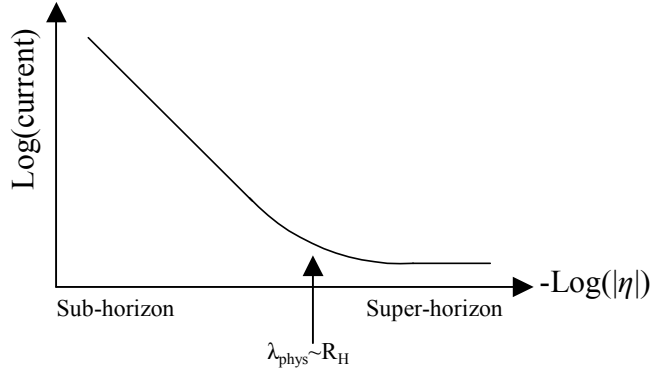

Figure(4.4) The current on with fixed $k$

## 5.3   The correlation function

We now analyze the energy fluctuations contributed by scalar fields $\phi$, which are present in addition to the inflaton field. We consider here a massless scalar field $\phi$ without self-interactions, minimally coupled to gravity.

Let us consider the time that just before the end of the de Sitter era. That is:

$$\eta = \eta_{RH} - \varepsilon .\qquad (5.51)$$

We remember that on (5.16) we have fixed the inflationary scale factor by setting $a_{RH} \equiv 1$. Therefore at this time the comoving and physical distances coincide and $k_{phys} = k$.

A key tool of [15] is that the computation of the correlation function $C(l) \equiv C(\vec{x}, \vec{x}')$ in a given quantum state $|\psi\rangle$ can be reduced to $\langle \psi | \delta\phi^\dagger(\vec{x}')\delta\phi(\vec{x}) | \psi \rangle$, Where $\delta\phi(\vec{x}) \equiv \phi(\vec{x}) - \phi_0$ and $\phi_0 \equiv \langle \phi(\vec{x}) \rangle$. We consider here that the field has no classical background component, i.e. with vanishing expectation values, $\langle \phi \rangle = 0$ [15]. So it is reduced to calculate the Wightman function of this state,

$$\begin{aligned}W(\vec{x}, \vec{x}') &\equiv \langle \psi | \phi^\dagger(\eta, \vec{x}')\phi(\eta, \vec{x}) | \psi \rangle \\ &= \int \frac{d^3k}{(2\pi)^3} u_k^*(\eta) u_k(\eta) e^{i\vec{k}(\vec{x}-\vec{x}')}\end{aligned}\qquad (5.52)$$

with the definition (5.23) $\varphi_k(\eta, \vec{x}) \equiv u_k(\eta) e^{i\vec{k}\vec{x}}$.

So we define the equal-time correlation function of two fields as follows,

$$C_\phi(l) \equiv \langle 0\ in|\phi^\dagger(\eta,\vec{x}')\phi(\eta,\vec{x})|0\ in\rangle$$
$$= \int \frac{d^3k}{(2\pi)^3} |u_k(\eta)|^2 e^{i\vec{k}\vec{l}} \quad . \tag{5.53}$$

Here the $l$ is independent of $\vec{x}$, therefore the correlation function is isotropic. And for $l \neq 0$ the correlation function is UV (ultraviolet) convergence.

With the result of (5.29) $\varphi_k(\eta,\vec{x}) = \frac{H}{\sqrt{2k^3}}(k\eta - i)e^{i(\vec{k}\vec{x}-k\eta)}$, for $m = 0$, so for a massless field the $|u_k(\eta)|^2$ is then

$$|u_k(\eta)|^2 = \left|\frac{H}{\sqrt{2k^3}}(k\eta - i)e^{-ik\eta}\right|^2 = \frac{H^2}{2k^3}\left(\sqrt{k^2\eta^2 + 1}\right)^2$$
$$= \frac{H^2}{2k^3}(k^2\eta^2 + 1) = \frac{1}{2k}\left(\frac{k^2\eta^2 H^2}{k^2} + \frac{H^2}{k^2}\right)$$
$$= \frac{1}{2k}\left(\eta^2 H^2 + \frac{H^2}{k^2}\right) \tag{5.54}$$
$$= \frac{1}{2k}\left(1 + \frac{H^2}{k^2}\right)\bigg|_{\eta=\eta_{RH}=-H^{-1}}$$

Therefore the correlation function for the massless fields is

$$C_\phi(l) = \int \frac{d^3k}{(2\pi)^3} \frac{1}{2k}\left(1 + \frac{H^2}{k^2}\right) e^{i\vec{k}\vec{l}} \quad . \tag{5.55}$$

We do not take into account modes with physical wavelengths larger than the Hubble radius today. Therefore the integral (5.55) must be cut off correspondingly at $k_{\min} = He^{-N}$ with $N \approx 70$ [15]. Thus we consider only $|\vec{k}| > k_{\min}$, this avoids the IR (infrared) divergence of the integral (5.55). And we obtain [15]

$$C_\phi(l) = \frac{1}{(2\pi)^2}\left[\frac{1}{l^2} - H^2 \ln\left(\frac{k_{\min}l}{4}\right)\right]. \tag{5.56}$$

The power spectrum of an operator $F$ is the Fourier transform of the 2-point correlation function $C_F(l)$,

$$P_F(k) = \int d^3l\, C_F(\vec{l}) e^{i\vec{k}\vec{l}} \tag{5.57}$$

and [8]

$$C_F(l) = \int \frac{d^3k}{(2\pi)^3} e^{i\vec{k}\vec{l}} P_F(k). \tag{5.58}$$

The integral (5.57) can be UV convergent or divergent. It depends on the short-distance behavior of the correlation function for the operator $F$. With

$$F(\vec{k}) = \int d^3x F(\vec{x}) e^{-i\vec{k}\vec{x}}. \tag{5.59}$$

One has [15]

$$\langle 0 \ in | F_{\vec{k}} F_{\vec{k}'}^\dagger | 0 \ in \rangle = (2\pi)^3 \delta^{(3)}(\vec{k} - \vec{k}') P_F(k). \tag{5.60}$$

For quantum field theory we take $F(\vec{x}) = \delta\phi(\vec{x})$. Because the expectation value vanishes, i.e. $\langle \phi \rangle = 0$, then is $F = \phi$. Therefore we obtain

$$P_\phi(k) = |u_k(\eta)|^2. \tag{5.61}$$

There is then no possibility of a UV or IR divergence in $P_\phi(k)$. From (5.54) we obtain for massless fields in de Sitter space-time:

$$P_\phi(k) = \frac{1}{2k}\left(1 + \frac{H^2}{k^2}\right). \tag{5.62}$$

We know that the free electric field $E$ in the Minkowski vacuum has the short-distance behavior $\langle 0 | E(\vec{x}) E(\vec{x}') | 0 \rangle \sim |\vec{x} - \vec{x}'|^{-4}$, and also $\langle 0 | E^2(\vec{x}) | 0 \rangle$, the mean square fluctuation at a given point, is infinite. So instead of consider $\langle 0 \ in | \phi^2(\vec{x}) | 0 \ in \rangle$, we must consider a field operator smeared in space (and time, in case of need.). Therefore the local operators must be smeared before squaring.[8]

$$F_R \equiv \int d^3x' W_R(|\vec{x} - \vec{x}'|) F(\vec{x}'). \tag{5.63}$$

Where $W_R$ is so called window function with the normalization

$$\int d^3x W_R(|\vec{x}|) = 1. \tag{5.64}$$

In our case we choose a gaussian window function

$$W_R(r) = \frac{e^{-r^2/2R^2}}{(2\pi R^2)^{3/2}}. \tag{5.65}$$

With $W_R(k) = \int d^3x e^{-i\vec{k}\vec{x}} W_R(|\vec{x}|)$ we obtain

$$W_R(k) = e^{\frac{-k^2 R^2}{2}}. \tag{5.66}$$

This is a gaussian UV cutoff. We notice that the normalization condition here is $W_R(k)|_{k=0} = 1$. And therefore $F_R(\vec{k}) = F(\vec{k}) W_R(k)$.

Because the mean square fluctuations of the smeared operator $\langle (F_R)^2 \rangle$ is independent of $\vec{x}$, so we can set $\vec{x} = 0$ and this becomes

$$\langle (F_R)^2 \rangle = \left\langle \int \frac{d^3k}{(2\pi)^3} F(\vec{k}) W_R(k) \int \frac{d^3k'}{(2\pi)^3} F(\vec{k}') W_R(k') \right\rangle$$
$$= \int \frac{d^3k}{(2\pi)^3} P_F(k) |W_R(k)|^2 \quad . \tag{5.67}$$

With $\int d^3k = \int_0^\infty dk \int_0^\pi d\vartheta \int_0^{2\pi} d\varphi\, k^2 \sin\vartheta = \int_0^\infty dk\, 4\pi k^2$ in spherical coordinates, the (5.67) will be then

$$\langle (F_R)^2 \rangle = \int_0^\infty \frac{dk}{(2\pi)^3} 4\pi k^2 P_F(k) |W_R(k)|^2$$
$$= \int_0^\infty \underbrace{\frac{dk}{k}}_{d(\ln k)} \left[ \frac{4\pi}{(2\pi)^3} k^3 P_F(k) \right] |W_R(k)|^2 \quad . \tag{5.68}$$
$$\equiv \int_0^\infty d(\ln k) \Delta_F^2(k) |W_R(k)|^2$$

With

$$\Delta_F^2(k) = \frac{k^3}{2\pi^2} P_F(k) . \tag{5.69}$$

So the $P_F(k)$ is then the power (fluctuation-strength) per unit volume $d^3k$, and the $\Delta_F^2(k)$ is the power per unit interval on the $\ln k$ axes, i.e. per e-fold in $k$.

With (5.62) $P_\phi(k) = \frac{1}{2k}\left(1 + \frac{H^2}{k^2}\right)$ for massless fields in de Sitter space-time, we have

$$\Delta_\phi^2(k) = \frac{k^2}{4\pi^2}\left(1 + \frac{H^2}{k^2}\right). \tag{5.70}$$

When on super-horizon limes $\lambda \gg R_H = \frac{1}{H}$, i.e. $k \ll H$ the second term is then dominated, and we now obtain:

$$\Delta_\phi^2(k) = \frac{H^2}{4\pi^2} . \tag{5.71}$$

This is the fundamental result from quantum field theory in de Sitter space-time used in the papers, which computed the density fluctuations in the inflationary universe [6].

We recall the current

$$\hat{j}_\mu = i : \phi^\dagger \overleftrightarrow{\partial}_\mu \phi : . \tag{5.72}$$

The hat refers to the normalization of the operator, so that for the 1-particle states $|\vec{k}\rangle = a^\dagger_{\vec{k}}|0\rangle$ is $\langle \vec{k}|\hat{j}^0|\vec{k}\rangle = 1/a^3$, i.e. 1 per unit comoving volume without the electric charge $e$ in front. The normal ordering of operators guarantees that $\langle 0 \ in|\hat{j}_\mu(\eta,\vec{x})|0 \ in\rangle = 0$.

Let us consider now the correlation function for $\vec{\nabla} \times \hat{j}(\vec{x})$, which is UV convergent for fixed separation between two points $\vec{x}$ and $\vec{x}'$, In Cartesian coordinates,

$$\left(\vec{\nabla} \times \hat{j}(\vec{x})\right)_i = \varepsilon_{ijk}\partial_j(-i\phi^\dagger \overleftrightarrow{\partial}_k \phi) = -2i\varepsilon_{ijk}(\partial_j\phi^\dagger)(\partial_k\phi), \quad i,j,k \in 1,2,3 . \tag{5.73}$$

The correlation function is evaluated by inserting a complete set of 2-particle states (here scalars fields $\pi^\pm$), this is

$$\begin{aligned} C_{\vec{\nabla}\times\hat{j}} &= \langle 0 \ in|(\vec{\nabla}\times\hat{j}_{\vec{x}})\cdot(\vec{\nabla}\times\hat{j}_{\vec{x}'})|0 \ in\rangle \\ &= -4\varepsilon_{ijk}\varepsilon_{ilm}\int \frac{d^3k}{(2\pi)^3}\int \frac{d^3k'}{(2\pi)^3}\langle 0 \ in|(\partial_j\phi^\dagger_{\vec{x}})(\partial_k\phi_{\vec{x}})|\pi^+_{\vec{k}}\pi^-_{\vec{k}'}\rangle\langle\pi^+_{\vec{k}}\pi^-_{\vec{k}'}|(\partial_m\phi^\dagger_{\vec{x}'})(\partial_n\phi_{\vec{x}'})|0 \ in\rangle \end{aligned} \tag{5.74}$$

The important observation of [15] is that such correlation functions can be written in terms of products of the elementary field correlation function $C_\phi(|\vec{x}-\vec{x}'|)$.

$$C_{\vec{\nabla}\times\hat{j}}(|\vec{x}-\vec{x}'|) = -4(\delta_{jl}\delta_{km} - \delta_{jm}\delta_{kl})\left(\frac{\partial}{\partial x^j}\frac{\partial}{\partial x'^m}C_\phi(|\vec{x}-\vec{x}'|)\right)\left(\frac{\partial}{\partial x^k}\frac{\partial}{\partial x'^l}C_\phi(|\vec{x}-\vec{x}'|)\right). \tag{5.75}$$

With use the relations,

$$\begin{aligned} \partial_{\vec{x}'}C_\phi(l) &= -\partial_{\vec{x}}C_\phi(l) \\ \partial_j\partial_m f(l) &= \hat{l}_j\hat{l}_m\frac{d^2 f}{dr^2} + (\delta_{jm} - \hat{l}_j\hat{l}_m)\left(\frac{1}{r}\frac{df}{dr}\right) \end{aligned} \tag{5.76}$$

Where $l \equiv |\vec{x}-\vec{x}'|$ and $\hat{l}_j \equiv l_j/l$. $(\delta_{jm} - \hat{l}_j\hat{l}_m) = (P_{trans})_{jm}$ is the transverse projector on the vector space orthogonal to $\hat{l}$, while $(\hat{l}_j\hat{l}_m) = (P_{long})_{jm}$ is the longitudinal projector on the direction $\hat{l}$. The longitudinal projectors do not contribute at all since $P_{trans} \cdot P_{long} = 0$ and the antisymmetry of $\delta_{jl}\delta_{km} - \delta_{jm}\delta_{kl}$. The $\delta_{jl}\delta_{km}$ contraction of the two transverse projectors gives $\text{trace}(P^2_{trans}) = 2$, while the $\delta_{jm}\delta_{kl}$ contraction gives $(\text{trace}\,P_{trans})^2 = 4$. Therefore

$$C_{\vec{\nabla}\times\hat{j}} = 8\left(\frac{1}{l}\frac{dC_\phi(l)}{dl}\right)^2 . \tag{5.77}$$

With (5.56) $C_\phi(l) = \frac{1}{(2\pi)^2}\left[\frac{1}{l^2} - H^2\ln\left(\frac{k_{\min}l}{4}\right)\right]$ for the massless fields, we obtain

$$C_{\bar{\nabla}\times\hat{j}} = \frac{8}{(2\pi)^4}\left(\frac{2}{l^4} + \frac{H^2}{l^2}\right)^2 = \frac{8H^8}{(2\pi)^4}\left(\frac{2}{H^4 l^4} + \frac{1}{H^2 l^2}\right)^2. \tag{5.78}$$

This is UV- and IR convergence, therefore there is no UV or IR cutoff. For physically relevant length scales at the end of inflation $l \gg H^{-1}$, only the second term contributes, that is

$$C_{\bar{\nabla}\times\hat{j}} = \frac{1}{2\pi^4}\frac{H^4}{l^4}. \tag{5.79}$$

# 6 The Evolution of the cosmological magnetic fields

Use the result of last chapter; we can now study the Evolution of the cosmological magnetic fields.

For a vector $\vec{V}$, it can be written as the sum of two terms,

$$\vec{V} = \vec{V}_l + \vec{V}_t \tag{6.1}$$

where $\vec{V}_l$ is called the longitudinal or irrotational current and has $\vec{\nabla} \times \vec{V}_l = 0$, while $\vec{V}_t$ is called the transverse or solenoidal component and has $\vec{\nabla} \cdot \vec{V}_t = 0$. Starting from the vector identity,

$$\vec{\nabla} \times (\vec{\nabla} \times \vec{V}) = \vec{\nabla}(\vec{\nabla} \cdot \vec{V}) - \Delta \vec{V} \tag{6.2}$$

together with

$$\Delta\left(\frac{1}{|\vec{x} - \vec{x}'|}\right) = -4\pi\delta(\vec{x} - \vec{x}'), \tag{6.3}$$

it can be shown that $\vec{V}_l$ and $\vec{V}_t$ can be constructed explicitly from $\vec{V}$ as follows [17]:

$$\vec{V}_l = -\frac{1}{4\pi}\vec{\nabla}\int d^3x' \frac{\vec{\nabla}' \cdot \vec{V}}{|\vec{x} - \vec{x}'|} \tag{6.4}$$

$$\vec{V}_t = \frac{1}{4\pi}\vec{\nabla} \times \vec{\nabla} \times \int d^3x' \frac{\vec{V}}{|\vec{x} - \vec{x}'|}. \tag{6.5}$$

We are interested in the divergence free transverse $\vec{B}$ component $\vec{B}_t$. Actually we may remember in the equation (4.19), there is $\vec{\nabla} \cdot \vec{B} = 0$.

Now we do a Fourier transformation for $\vec{B}$,

$$\vec{B}(\vec{x},\eta) = \int \frac{d^3k}{(2\pi)^3} \vec{B}(\vec{k},\eta)e^{i\vec{k}\vec{x}}. \tag{6.6}$$

Where $\vec{B}(\vec{k},\eta) \equiv \vec{b}(\vec{k},\eta)e^{-i\omega\eta}$ denotes the amplitude for $\vec{k}$.

Here we do also a Fourier transformation for $\vec{j}$:

$$\vec{j}(\vec{x},\eta) = \int \frac{d^3k}{(2\pi)^3} \vec{j}(\vec{k},\eta)e^{i\vec{k}\vec{x}}. \tag{6.7}$$

We consider now a single mode $\vec{B}(\vec{k},\eta)e^{i\vec{k}\vec{x}}$, and put it into the wave equation (4.36),

$$\Delta\left(\vec{B}(\vec{k},\eta)e^{i\vec{k}\vec{x}}\right) - \frac{1}{a^4}\frac{\partial^2\left(\vec{B}(\vec{k},\eta)e^{i\vec{k}\vec{x}}\right)}{\partial\eta^2} + \text{curl}\left(\vec{j}(\vec{k},\eta)e^{i\vec{k}\vec{x}}\right) = 0. \tag{6.8}$$

From appendix A we see

$$\Delta S = \frac{1}{a^2}\left(\frac{\partial}{\partial x^1}\frac{\partial S}{\partial x^1} + \frac{\partial}{\partial x^2}\frac{\partial S}{\partial x^2} + \frac{\partial}{\partial x^3}\frac{\partial S}{\partial x^3}\right). \tag{6.9}$$

Then we get

$$-\frac{k^2}{a^2}\vec{B}(\vec{k},\eta) - \frac{1}{a^4}\frac{d^2\left(a^2\vec{B}(\vec{k},\eta)\right)}{d\eta^2} + \frac{1}{a}i\vec{k}\times\vec{j}(\vec{k},\eta) = 0. \tag{6.10}$$

This is

$$a^2k^2\vec{B}(\vec{k},\eta) + \frac{d^2\left(a^2\vec{B}(\vec{k},\eta)\right)}{d\eta^2} - ia^3\vec{k}\times\vec{j}(\vec{k},\eta) = 0. \tag{6.11}$$

We now define

$$\begin{aligned}\vec{b} &\equiv a^2\vec{B}(\vec{k},\eta) \\ \vec{s} &\equiv a^3\vec{k}\times\vec{j}(\vec{k},\eta)\end{aligned} \tag{6.12}$$

Thus we obtain a linear differential equation

$$\frac{d^2\vec{b}}{d\eta^2} + k^2\vec{b} = \vec{s}. \tag{6.13}$$

Before we go on, we do the following 2 assumptions:

1.) The inflation lasts for $N_e$ e-folds. It means:

$$\frac{a_{RH}}{a_{ini}} \equiv e^{N_e}. \tag{6.14}$$

Where $a_{ini}$ denotes at the beginning of the inflation.

2.) On the earlier time (ad hoc)

$$\text{source} = 0. \tag{6.15}$$

With the equation (6.14) $\frac{a_{RH}}{a_{ini}} \equiv e^{N_e}$, and the time on reheating (5.16) $\eta_{RH} = -\frac{1}{H}$, we obtain

$$\eta_{ini} = -\frac{e^{N_e}}{H}. \tag{6.16}$$

Now let us consider the homogeneous equation

$$\frac{d^2\vec{b}}{d\eta^2} + k^2\vec{b} = 0 \,. \tag{6.17}$$

We have the general solutions

$$\vec{b}(\eta) = \vec{c}e^{ik\eta} + \vec{d}e^{-ik\eta} \qquad (\vec{c}, \vec{d} \text{ denote complex const vector.}) \,. \tag{6.18}$$

If we know the boundary conditions:

$$\begin{aligned} \vec{b}(0) &= \vec{A} \\ \left. \frac{d\vec{b}(\eta)}{d\eta} \right|_{\eta=0} &= \vec{B} \end{aligned} \tag{6.19}$$

Then we can calculate the $\vec{c}, \vec{d}$, and we obtain

$$\vec{b}(\eta) = \vec{A}\cos k\eta + \vec{B}\frac{1}{k}\sin k\eta \,. \tag{6.20}$$

We notice here, we used the formulae $\sin x = \frac{1}{2i}(e^{ix} - e^{-ix})$ and $\cos x = \frac{1}{2}(e^{ix} + e^{-ix})$.

Let us consider the situation with the source $\vec{s}$.

At first one solves the Green function for this equation,

$$\frac{d^2 g(\eta;\eta')}{d\eta^2} + k^2 g(\eta;\eta') = \delta(\eta - \eta'), \quad \eta_{ini} < \eta < \eta_{RH} \,. \tag{6.21}$$

With the boundary condition for our retarded Green function

$$\begin{aligned} g^{ret}(\eta';\eta') &= 0 \\ \left. \frac{dg^{ret}(\eta;\eta')}{d\eta} \right|_{\eta=\eta'} &= 1 \end{aligned} \tag{6.22}$$

We then obtain

$$g^{ret}(\eta;\eta') = \frac{1}{k}\sin k(\eta - \eta') \,. \tag{6.23}$$

And the inhomogeneous solution $b(\eta)$ is:

$$\vec{b}(\eta) = \int_{\eta_{ini}}^{\eta_{RH}} g(\eta;\eta')\vec{s}(\eta')d\eta' \,. \tag{6.24}$$

Recall (5.46) and we get

$$\frac{d^2\vec{b}(\eta)}{d\eta^2} + K^2\vec{b} = ia^4 \left\langle \pi_{\vec{k}_+}^+, \pi_{\vec{k}_-}^- \left| \frac{1}{a}\vec{K} \times \vec{j}(\vec{k},\eta) \right| 0 \ in \right\rangle$$
$$= ie\frac{H^2 a^2}{\sqrt{2k_+^3}\sqrt{2k_-^3}}(k_+\eta - i)(k_-\eta - i)(\vec{k}_+ + \vec{k}_-) \times (\vec{k}_+ - \vec{k}_-)e^{i(\omega_+ + \omega_-)\eta} \ . \quad (6.25)$$
$$(\vec{K} \equiv \vec{k}_+ + \vec{k}_-, \ \omega \equiv \omega_+ + \omega_- = k_+ + k_-)$$

We recall (5.21), in de Sitter space-time

$$a(\eta) = -\frac{1}{H\eta} \ . \quad (6.26)$$

Then we obtain

$$\frac{d^2\vec{b}(\eta)}{d\eta^2} + K^2\vec{b}$$
$$= ie\frac{H^2}{H^2\eta^2\sqrt{2k_+^3}\sqrt{2k_-^3}}\left(k_+k_-\eta^2 - i(k_+ + k_-)\eta - 1\right)(\vec{k}_+ + \vec{k}_-) \times (\vec{k}_+ - \vec{k}_-)e^{i(\omega_+ + \omega_-)\eta} \ . \quad (6.27)$$
$$= ie\frac{1}{\eta^2\sqrt{2k_+^3}\sqrt{2k_-^3}}\left(k_+k_-\eta^2 - 1 - i(k_+ + k_-)\eta\right)(\vec{k}_+ + \vec{k}_-) \times (\vec{k}_+ - \vec{k}_-)e^{i(\omega_+ + \omega_-)\eta}$$

So from (6.24) we obtain

$$\vec{b}(\eta) = \int_{\eta_{ini}}^{\eta_{RH}} \frac{1}{K}\sin\left(K(\eta - \eta')\right) \cdot ie \frac{1}{\eta'^2\sqrt{2k_+^3}\sqrt{2k_-^3}}\left(k_+k_-\eta'^2 - 1 - i(k_+ + k_-)\eta'\right)(\vec{k}_+ + \vec{k}_-) \times (\vec{k}_+ - \vec{k}_-)e^{i(k_+ + k_-)\eta'}d\eta'$$
$$. \quad (6.28)$$

We know

$$k = \frac{2\pi}{\lambda}$$
$$\omega = 2\pi\nu = 2\pi\frac{c}{\lambda} = k, \qquad c = 1 \quad (6.29)$$

For our convenience we suppose

$$\vec{k}_+ = k\vec{e}_+$$
$$\vec{k}_- = k\vec{e}_-, \quad \vec{e}_+ \perp \vec{e}_- \quad (6.30)$$

and we obtain

$$\vec{b}(\eta) = ie\int_{\eta_{ini}}^{\eta_{RH}} \frac{1}{\sqrt{2}k}\sin\left(\sqrt{2}k(\eta - \eta')\right) \cdot \frac{1}{\eta'^2 2k^3}\left(k^2\eta'^2 - 1 - 2ik\eta'\right)2k^2\vec{e}_3 e^{2ik\eta'}d\eta' \ . \quad (6.31)$$

Where $\vec{e}_3$ denotes the direction that orthogonal to $\vec{e}_+$ and $\vec{e}_-$. So we define

$$b(\eta) \equiv i\frac{e}{\sqrt{2}} \int_{\eta_{ini}}^{\eta_{RH}} \sin\left(\sqrt{2}k(\eta - \eta')\right) \cdot \left(1 - \frac{1}{k^2\eta'^2} - 2i\frac{1}{k\eta'}\right) e^{2ik\eta'} d\eta'$$

$$= i\frac{e}{\sqrt{2}} \underbrace{\int_{\eta_{ini}}^{\eta_{RH}} \sin\left(\sqrt{2}k(\eta - \eta')\right) \cdot e^{2ik\eta'} d\eta'}_{\equiv b_A} + i\frac{e}{\sqrt{2}} \underbrace{\int_{\eta_{ini}}^{\eta_{RH}} \sin\left(\sqrt{2}k(\eta - \eta')\right) \cdot \left(-\frac{1}{k^2\eta'^2} - 2i\frac{1}{k\eta'}\right) e^{2ik\eta'} d\eta'}_{\equiv b_B}$$

$$\equiv b_A + b_B \tag{6.32}$$

With the formula

$$\int dx e^{ax} \sin bx = \frac{e^{ax}}{a^2 + b^2}(a \sin bx - b \cos bx) \tag{6.33}$$

we obtain

$$b_A = i\frac{e}{\sqrt{2}} \frac{e^{2ik\eta'}}{-2k}\left(i2\sin\sqrt{2}k(\eta - \eta') + \sqrt{2}\cos\sqrt{2}k(\eta - \eta')\right)\Bigg|_{\eta_{ini}}^{\eta_{RH}}. \tag{6.34}$$

Since we suppose at earlier time $\eta_{ini}$ source $= 0$ (6.15), and for our convenience we suppose also at the time $\eta_{RH}$ the phase is equal to zero then $e^{ik\eta_{RH}} = 1$. This is justifiable since we are only interested on the large-scale cosmologic magnetic fields and the $\eta_{RH}$ is fixed by (5.16) $\eta_{RH} \equiv -H^{-1}$. Then the (6.34) will be

$$b_A = \frac{e}{2\sqrt{2}k}\left(2\sin\sqrt{2}k\eta - i\sqrt{2}\cos\sqrt{2}k\eta\right). \tag{6.35}$$

With $\sin x = \frac{1}{2i}(e^{ix} - e^{-ix})$ and the partial integral

$$\int_{x_0}^{x_1} dx' \frac{e^{iax'}}{x'^2} = -\frac{e^{iax'}}{x'}\Bigg|_{x_0}^{x_1} + ia\int_{x_0}^{x_1} dx' \frac{1}{x'} e^{iax'} \tag{6.36}$$

we obtain

$$b_B = -\frac{ieH}{\sqrt{2}k^2}\sin\sqrt{2}k\eta - i\frac{e(4-\sqrt{2})}{2\sqrt{2}k}e^{i\sqrt{2}k\eta}\int_{k\eta_{ini}}^{k\eta_{RH}} \frac{e^{i(2-\sqrt{2})z'}}{z'}dz' + i\frac{e(4+\sqrt{2})}{2\sqrt{2}k}e^{-i\sqrt{2}k\eta}\int_{k\eta_{ini}}^{k\eta_{RH}} \frac{e^{i(2+\sqrt{2})z'}}{z'}dz' .$$

$$\tag{6.37}$$

We notice here $\eta_{RH} = -\frac{1}{H}$. Since

$$\int\limits_{k\eta_{ini}}^{k\eta_{RH}} \frac{e^{i(2-\sqrt{2})z'}}{z'} dz' \leq \left| \int\limits_{k\eta_{ini}}^{k\eta_{RH}} \frac{e^{i(2-\sqrt{2})z'}}{z'} dz' \right| \leq \ln(-k\eta_{RH}) = \ln\frac{k}{H}. \tag{6.38}$$

This is logarithmic; in comparison to other terms this term is negligible. And we obtain

$$b_B \approx -\frac{ieH}{\sqrt{2}k^2} \sin\sqrt{2}k\eta. \tag{6.39}$$

So

$$b = b_A + b_B \approx \frac{e}{2\sqrt{2}k}\left(2\sin\sqrt{2}k\eta - i\sqrt{2}\cos\sqrt{2}k\eta\right) - \frac{ieH}{\sqrt{2}k^2}\sin\sqrt{2}k\eta. \tag{6.40}$$

Since our interest is the large-scale cosmological magnetic fields, this means $k \ll 1$ and the first term is therefore negligible and we obtain

$$|b(\eta)| \approx \frac{eH}{\sqrt{2}k^2}\left|\sin\sqrt{2}k\eta\right|. \tag{6.41}$$

We see at first when we don't consider the oscillation of the $b(\eta)$, then it is conserved. With (6.12) $b = aB^2$, we see

$$b = a^2 B \approx \frac{eH}{\sqrt{2}k^2}. \tag{6.42}$$

So our conclusion is that during the inflation the $\vec{B}$ field is strongly reduced with factor $1/a^2$, and because of the $1/k^2$ factor the low frequency $B$ is much bigger than the high frequency.

Afterwards let us consider just before the end of de Sitter era, i.e., (5.51) $\eta = \eta_{RH} - \varepsilon$. With $a_{RH} \equiv 1$ one gets $\vec{k}_{phys} \approx \vec{k}$. We then obtain

$$B \approx \frac{eH}{\sqrt{2}k^2}\sqrt{2}k|\eta| = \frac{eH}{k_{phys}}|\eta|. \tag{6.43}$$

We notice $\sin x \approx x$, for $|x| \ll 1$. With (5.16) $\eta_{RH} \equiv -H^{-1}$ one gets at the time $\eta_{RH}$

$$B_{RH} = \frac{eH}{k_{phys}}\frac{1}{H} = \frac{e}{k_{phys}}. \tag{6.44}$$

# 7      Acknowledgments

I would like to thank my supervisor Prof. Christoph Schmid for his guidance and time. I would like to express my thanks to Jan Michael Kratochvil of Stanford University for the discussion of one integral calculation. I am also grateful to Dr. Andrew Matterson and Dr. Ting-Lin Shang for the English corrections.

# 8 Appendix A: Gradient, Curl, and Divergence

Generally we use covariant derivatives in a curved space-time. Which form will they take? Simplest of all is the covariant derivative of a scalar, which is just the ordinary gradient

$$S_{;\mu} = \frac{\partial S}{\partial x^\mu}. \tag{8.1}$$

We know that

$$V_{\mu;\nu} \equiv \frac{\partial V_\mu}{\partial x^\nu} - \Gamma^\lambda{}_{\mu\nu} V_\lambda$$
$$V^\mu{}_{;\nu} \equiv \frac{\partial V^\mu}{\partial x^\nu} + \Gamma^\mu{}_{\nu\lambda} V^\lambda \tag{8.2}$$

Since $\Gamma^\lambda{}_{\mu\nu}$ is symmetric in $\mu$ and $\nu$, the covariant curl is just the ordinary curl

$$V_{\mu;\nu} - V_{\nu;\mu} = \frac{\partial V_\mu}{\partial x^\nu} - \frac{\partial V_\nu}{\partial x^\mu}. \tag{8.3}$$

Another special case is the covariant divergence of a contravariant vector

$$V^\mu{}_{;\mu} \equiv \frac{\partial V^\mu}{\partial x^\mu} + \Gamma^\mu{}_{\mu\lambda} V^\lambda. \tag{8.4}$$

We define now

$$g \equiv \det g_{\mu\nu}. \tag{8.5}$$

One notices, we assume we are in a time-like world, so

$$\det g_{\mu\nu} < 0. \tag{8.6}$$

After calculation [12], we get

$$V^\mu{}_{;\mu} = \frac{1}{\sqrt{-g}} \frac{\partial}{\partial x^\mu} \sqrt{-g} V^\mu. \tag{8.7}$$

Now we want to know what the tensor analysis formalism outlined above has to do with the familiar formulas for gradient, curl and divergence.

At first we define our metric as follows [14]

$$g_{ij} \equiv \vec{g}_i \cdot \vec{g}_j = g_{ji} \quad i,j = 1,2,3. \tag{8.8}$$

Where covariant vector is defined as

$$\vec{g}_i \equiv \frac{\partial \vec{r}}{\partial x^i}. \tag{8.9}$$

And the contravariant definition is

$$g^{ij} \equiv \vec{g}^i \cdot \vec{g}^j = g^{ji}. \tag{8.10}$$

Where

$$\vec{g}^i \equiv \nabla x^i. \tag{8.11}$$

Some features are as follows [14]

$$\vec{g}_i \cdot \vec{g}^k = \frac{\partial \vec{r}}{\partial x^i} \cdot \vec{\nabla} x^k = \frac{\partial x^k}{\partial x^i} = \delta_i^k, \tag{8.12}$$

$$\sum_i g_{ik} g^{il} = \delta_k^l. \tag{8.13}$$

Vectors can be written down as following

$$\begin{aligned}\vec{a} &= \sum_i \vec{g}_i (\vec{g}^i \cdot \vec{a}) \equiv \sum_i \vec{g}_i a^i \\ \vec{a} &= \sum_i \vec{g}^i (\vec{g}_i \cdot \vec{a}) \equiv \sum_i \vec{g}^i a_i\end{aligned}. \tag{8.14}$$

Just like (8.5), in 3-dimension we define

$$g \equiv \det g_{ij}. \tag{8.15}$$

In this paper we use the convention

$$\eta_{\mu\nu} = \begin{pmatrix} 1 & & & \\ & -1 & & \\ & & -1 & \\ & & & -1 \end{pmatrix}. \tag{8.16}$$

So in general coordinates we have the following formulas [14]

$$\operatorname{grad} \psi \equiv \vec{\nabla} \psi = \sum_i \vec{g}^i (\vec{g}_i \cdot \vec{\nabla} \psi) = \sum_i \vec{g}^i \frac{\partial \psi}{\partial x^i}, \tag{8.17}$$

$$\operatorname{div} \vec{a} \equiv \vec{\nabla} \cdot \vec{a} = \frac{1}{\sqrt{-g}} \sum \frac{\partial}{\partial x^i} (\sqrt{-g} a^i), \tag{8.18}$$

$$\operatorname{curl} \vec{a} \equiv \vec{\nabla} \times \vec{a} = \sum_{ikl} \varepsilon^{ikl} \vec{g}_i \frac{\partial a_l}{\partial x^k}. \tag{8.19}$$

Now let us consider a three-dimensional orthonormal coordinate system characterized by the metric

Cosmological Magnetic Fields: generation during Inflation and Evolution                                - 38 -

$$g_{ij} = \pm h_i^2 \delta_{ij} \quad (i,j = 1,2,3). \tag{8.20}$$

We can define the orthonormal basis

$$\vec{e}_i = \frac{\vec{g}_i}{h_i} = h_i \vec{g}^{\,i}. \tag{8.21}$$

So we can see

$$d\vec{r} = \sum_i \vec{e}_i h_i dx^i = \frac{\vec{g}_i}{h_i} = \vec{g}^{\,i} h_i, \tag{8.22}$$

$$\begin{aligned}\sqrt{-g} &= h_1 h_2 h_3, \text{ for } g_{ij} = -h_i^2 \delta_{ij} \\ \sqrt{g} &= h_1 h_2 h_3, \text{ for } g_{ij} = h_i^2 \delta_{ij}\end{aligned} \tag{8.23}$$

And

$$\begin{aligned}a_i &= (\vec{a}\cdot\vec{e}_i) h_i \\ a^i &= (\vec{a}\cdot\vec{e}_i)/h_i\end{aligned} \tag{8.24}$$

Now we get

$$\operatorname{grad}\psi \equiv \vec{\nabla}\psi = \sum_i \vec{e}_i \frac{1}{h_i}\frac{\partial \psi}{\partial x^i}, \tag{8.25}$$

$$\operatorname{div}\vec{a} \equiv \vec{\nabla}\cdot\vec{a} = \frac{1}{h_1 h_2 h_3}\sum_i \frac{\partial h_1 h_2 h_3 a^i}{\partial x^i}, \tag{8.26}$$

$$\vec{e}_1 \cdot (\operatorname{curl}\vec{a}) \equiv \vec{e}_1 \cdot (\vec{\nabla}\times\vec{a}) = \frac{1}{h_2 h_3}\left(\frac{\partial a_3}{\partial x^2} - \frac{\partial a_2}{\partial x^3}\right) \quad \text{(and cyclical)}. \tag{8.27}$$

What are usually called the components of a vector $\vec{a}$ in elementary treatments are not the covariant components $a_i$ or the contravariant components $a^i$, but the "ordinary" components [12]

$$\bar{a}_i \equiv \vec{a}\cdot\vec{e}_i = h_i a^i = h_i^{-1} a_i. \tag{8.28}$$

Then we can rewrite the equations (8.26) – (8.27) in "ordinary" components

$$\operatorname{div}\vec{a} \equiv \vec{\nabla}\cdot\vec{a} = \frac{1}{h_1 h_2 h_3}\left(\frac{\partial h_2 h_3 \bar{a}_1}{\partial x^1} + \frac{\partial h_1 h_3 \bar{a}_2}{\partial x^2} + \frac{\partial h_1 h_2 \bar{a}_3}{\partial x^3}\right), \tag{8.29}$$

$$\vec{e}_1 \cdot (\operatorname{curl}\vec{a}) \equiv \vec{e}_1 \cdot (\vec{\nabla}\times\vec{a}) = \frac{1}{h_2 h_3}\left(\frac{\partial h_3 \bar{a}_3}{\partial x^2} - \frac{\partial h_2 \bar{a}_2}{\partial x^3}\right) \quad \text{(and cyclical)}. \tag{8.30}$$

Let us do an exercise to show how it works. We consider a sphere coordinate $r, \theta, \varphi$ with local orthonormal basis $\vec{e}_r, \vec{e}_\theta, \vec{e}_\varphi$. We know

$$\vec{g}_1 = \vec{e}_r = \vec{r}/r$$
$$\vec{g}_2 = r\vec{e}_\theta \qquad . \qquad (8.31)$$
$$\vec{g}_3 = r\sin\theta \vec{e}_\varphi$$

And we get

$$d\vec{r} = \vec{e}_r dr + \vec{e}_\theta r d\theta + \vec{e}_\varphi r \sin\theta d\varphi. \qquad (8.32)$$

The metric is

$$g_{ij} = \begin{pmatrix} 1 & & \\ & r^2 & \\ & & r^2 \sin^2\theta \end{pmatrix}. \qquad (8.33)$$

Then we get the explicit form in sphere coordinate:

$$\operatorname{grad}\psi \equiv \vec{\nabla}\psi = \vec{e}_r \frac{\partial \psi}{\partial r} + \vec{e}_\theta \frac{1}{r}\frac{\partial \psi}{\partial \theta} + \vec{e}_\varphi \frac{1}{r\sin\theta}\frac{\partial \psi}{\partial \varphi}, \qquad (8.34)$$

$$\operatorname{div}\vec{a} \equiv \vec{\nabla}\cdot\vec{a} = \frac{1}{r^2}\frac{\partial a_r r^2}{\partial r} + \frac{1}{r\sin\theta}\frac{\partial a_\theta \sin\theta}{\partial \theta} + \frac{1}{r\sin\theta}\frac{\partial a_\varphi}{\partial \varphi}, \qquad (8.35)$$

$$\operatorname{curl}\vec{a} \equiv \vec{\nabla}\times\vec{a} = \vec{e}_r \frac{1}{r\sin\theta}(\frac{\partial a_\varphi \sin\theta}{\partial \theta} - \frac{\partial a_\theta}{\partial \varphi}) + \vec{e}_\theta \frac{1}{r}(\frac{1}{\sin\theta}\frac{\partial a_r}{\partial \varphi} - \frac{\partial r a_\varphi}{\partial r}) + \vec{e}_\varphi \frac{1}{r}(\frac{\partial r a_\theta}{\partial r} - \frac{\partial a_r}{\partial \theta}) \qquad .$$
$$(8.36)$$

We notice here, $a_r, a_\theta, a_\varphi$ are the "ordinary" components $\bar{a}_i \equiv \vec{a}\cdot\vec{e}_i$.

## 9 Appendix B: The wave equation of the electric fields in FRW spatially flat universe

We write down again the Maxwell equation (4.19) and (4.20) here,

$$\frac{\partial}{\partial t} a^2 \vec{B} + a^2 \operatorname{curl} \vec{E} = 0,$$
$$\operatorname{div} \vec{B} = 0$$
(9.1)

$$\operatorname{curl} \vec{B} - \frac{1}{a^2} \frac{\partial}{\partial t} a^2 \vec{E} = \vec{j}.$$
$$\operatorname{div} \vec{E} = \rho$$
(9.2)

From (9.1) we get

$$\operatorname{curl} \vec{E} = -\frac{1}{a^2} \frac{\partial}{\partial t} a^2 \vec{B}.$$
(9.3)

We then use the vector identity $\operatorname{curl}\operatorname{curl} \vec{v} = \operatorname{grad}(\operatorname{div} \vec{v}) - \Delta \vec{v}$ and we know $\operatorname{div} \vec{E} = \rho$. Here we get

$$\operatorname{grad}(\operatorname{div} \vec{E}) - \Delta \vec{E} = \operatorname{curl}\left(-\frac{1}{a^2} \frac{\partial}{\partial t} a^2 \vec{B}\right)$$
$$\Rightarrow \operatorname{grad} \rho - \Delta \vec{E} = -\frac{1}{a^2} \operatorname{curl}\left(\frac{\partial}{\partial t} a^2 \vec{B}\right).$$
(9.4)

We define here

$$\dot{a} \equiv \frac{\partial a}{\partial t}.$$
(9.5)

We recall (4.28)

$$\left([\operatorname{curl}, \frac{\partial}{\partial t}]\vec{V}\right)_1 = \left(\operatorname{curl}\left(\frac{\partial}{\partial t} \vec{V}\right) - \frac{\partial}{\partial t}\left(\operatorname{curl} \vec{V}\right)\right)_1$$
$$= \frac{\dot{a}}{a}\left(\operatorname{curl} \vec{V}\right)_1.$$
(9.6)

So we get

$$\operatorname{grad} \rho - \Delta \vec{E} = -\frac{1}{a^2}\left(\frac{\partial}{\partial t}\left(\operatorname{curl} a^2 \vec{B}\right) + \frac{\dot{a}}{a} \operatorname{curl} a^2 \vec{B}\right).$$
(9.7)

From (9.2) we know:

$$\operatorname{curl} a^2 \vec{B} = a^2 \operatorname{curl} \vec{B} = \frac{\partial}{\partial t}\left(a^2 \vec{E}\right) + a^2 \vec{j}.$$
(9.8)

Now we put the equation (9.8) to (9.7)

$$\operatorname{grad}\rho - \Delta\vec{E} = -\frac{1}{a^2}\left(\frac{\partial}{\partial t}\left(\frac{\partial}{\partial t}\left(a^2\vec{E}\right) + a^2\vec{j}\right) + \frac{\dot{a}}{a}\left(\frac{\partial}{\partial t}\left(a^2\vec{E}\right) + a^2\vec{j}\right)\right)$$
$$= -\frac{1}{a^2}\frac{\partial^2}{\partial t^2}\left(a^2\vec{E}\right) - \frac{1}{a^2}\frac{\partial}{\partial t}\left(a^2\vec{j}\right) - \frac{\dot{a}}{a^3}\frac{\partial}{\partial t}\left(a^2\vec{E}\right) - \frac{\dot{a}}{a}\vec{j} \quad (9.9)$$

And we know

$$H \equiv \frac{\partial a}{\partial t}\Big/a = \frac{\dot{a}}{a}. \quad (9.10)$$

So now we put this into equation (9.9) and get

$$\Delta\vec{E} - \frac{1}{a^2}\frac{\partial^2\left(a^2\vec{E}\right)}{\partial t^2} - \frac{H}{a^2}\frac{\partial\left(a^2\vec{E}\right)}{\partial t} - \frac{1}{a^2}\frac{\partial\left(a^2\vec{j}\right)}{\partial t} - H\vec{j} - \operatorname{grad}\rho = 0. \quad (9.11)$$

And this is the wave equation of electric Field.

With conformal time $\eta$, we have

$$\Delta\vec{E} - \frac{1}{a^4}\frac{\partial^2\left(a^2\vec{E}\right)}{\partial\eta^2} - \frac{1}{a^3}\frac{\partial\left(a^2\vec{j}\right)}{\partial\eta} - H\vec{j} - \operatorname{grad}\rho = 0. \quad (9.12)$$